\documentclass[11pt,a4paper]{article}
\usepackage{jcappub}
\usepackage{physics}

\usepackage{amssymb,amsfonts}
\usepackage{amssymb} 
\usepackage{amsmath, amsthm} 
\usepackage{amsgen, amsfonts, amsbsy} 
\usepackage{mathrsfs} 
\usepackage{color}
\newcommand{\be}{\begin{equation}}
\newcommand{\ee}{\end{equation}}

\def\theequation{\arabic{section}.\arabic{equation}}

\title{\textbf{Disformal mappings of spherical DHOST geometries}}

\author[a]{Valerio Faraoni,}
\author[b]{Alexandre Leblanc}

\affiliation[a]{Department of Physics \& Astronomy, Bishop's University,\\
2600 College Street, Sherbrooke, Quebec, Canada}
\affiliation[b]{Department of Physics, Universit\'e de Sherbrooke,\\
2500 Boulevard de l'Universit\'e, Sherbrooke, Quebec, Canada}

\emailAdd{vfaraoni@ubishops.ca}
\emailAdd{Alexandre.Leblanc3@USherbrooke.ca}

\abstract{New solutions of DHOST theories can be generated by applying a 
disformal tranformation to a known seed solution. We examine the nature of 
spherically symmetric solutions of DHOST gravity obtained by disforming 
static spherical scalar field solutions, or stealth solutions, of general 
relativity. It is shown that, in these cases, black hole horizons are 
never created by disforming a black hole seed. New DHOST solutions are 
then created by disforming two lesser known scalar field solutions of 
general relativity: Wyman's ``other'' solution and the 
Husain-Martinez-Nu\~nez one. These new solutions demonstrate that one can 
obtain black hole horizons, wormhole throats, or horizonless geometries by 
disforming non-stealth, time-dependent, seeds.}


\arxivnumber{}

\begin{document}
\maketitle
\flushbottom

\section{Introduction}
\label{sec:1}
\setcounter{equation}{0}

The theoretical need to explain the present acceleration of the cosmic 
expansion without invoking an {\em ad hoc} dark energy 
\cite{AmendolaTsujikawabook} keeps stimulating the study of modifications 
of gravity with respect to Einstein's general relativity (GR). It is quite 
possible that the observed cosmic acceleration is the manifestation of 
deviations from GR at large scales. One of the most popular candidates for 
modified gravity is the $f({\cal R})$ family of theories (where ${\cal R}$ 
is the Ricci scalar), a subclass of scalar-tensor gravity 
\cite{Brans:1961sx, Bergmann:1968ve, Nordtvedt:1968qs, Wagoner:1970vr, 
Nordtvedt:1970uv} (see \cite{Sotiriou:2008rp, 
DeFelice:2010aj,Nojiri:2010wj} for reviews). However, new scalar-tensor 
theories of gravity (in addition to the original Brans-Dicke theory 
\cite{Brans:1961sx} and its ``first generation'' generalizations 
\cite{Bergmann:1968ve, Nordtvedt:1968qs, Wagoner:1970vr, 
Nordtvedt:1970uv}) have emerged and have been the subject of intense study 
for almost a decade, beginning with the rediscovery and reformulation 
\cite{H1,H2,H3} of Horndeski gravity \cite{Horndeski}. The field equations 
of Horndeski gravity are of second order, thus avoiding the notorius 
Ostrogradski instability affecting higher order theories. Surprisingly, 
certain theories with higher order equations of 
motion reduce, under the imposition of certain degeneracy conditions, to 
healthy theories in this regard \cite{GLPV1,GLPV2}. These Degenerate 
Higher Order Scalar-Tensor (or ``DHOST'') theories were developed in 
\cite{DHOST1, DHOST3, DHOST3, DHOST4, DHOST5, DHOST6, DHOST7} and have 
generated a rich literature (see Refs.~\cite{DHOSTreview1, DHOSTreview2} 
for reviews). The recent multi-messenger observation of a simultaneous 
gravitational wave and gamma-ray burst GW170817/GRB170817 
\cite{TheLIGOScientific:2017qsa,Monitor:2017mdv} sets severe constraints 
on the space of DHOST theories by restricting the difference between the 
propagation speeds of gravitational and electromagnetic waves 
\cite{Langloisetal18}. Theoretical constraints avoiding graviton decay 
into scalar field perturbations further restrict the space of allowable 
theories \cite{Creminellietal18}.

We follow the notation of Ref.~\cite{Wald:1984rg}: the metric signature is 
${-}{+}{+}{+}$ and units are used in which the speed of light and Newton's 
constant are unity. In this notation, the general DHOST action is
\begin{eqnarray}
S_{(\text{DHOST})} \left[ g_{ab}, \phi \right] &=& \int d^4 x \, \sqrt{-g} 
\left.\Big\{f_0\left( \phi, \bar{X}\right) +f_1\left( \phi, \bar{X} \right) 
\Box\phi 
+ f_2\left( \phi, \bar{X} \right) {\cal R} \right.\nonumber\\
&&\nonumber\\
&\, & \left. + {\cal A}_{(2)}^{abcd}\nabla_a\nabla_b\phi  
\nabla_c\nabla_d\phi
+f_3\left( \phi, \bar{X} \right) G_{ab} \nabla^a\nabla^b\phi 
\right.\nonumber\\
&&\nonumber\\
&\, & \left. + {\cal A}_{(3)}^{abcdef}\nabla_a\nabla_b\phi  
\nabla_c\nabla_d\phi 
\nabla_e\nabla_f \phi \right\} \,, \label{DHOSTaction}
\end{eqnarray}
where $\phi$ is the scalar field degree of freedom, $X \equiv \nabla^c 
\phi \nabla_c \phi$, $\nabla_a $ is the covariant derivative of the 
metric $g_{ab}$ (which has determinant $g$), and $\Box \equiv g^{ab} 
\nabla_a \nabla_b $ is d'Alembert's operator, while $G_{ab}$ denotes the 
Einstein tensor. 

The quadratic terms in the action~(\ref{DHOSTaction}) are usually written 
as 
\be
{\cal A}_{(2)}^{abcd} \nabla_a\nabla_b \phi \nabla_c\nabla_d \phi  
=\sum_{i=1}^5 \, \alpha_i \left( \phi, \bar{X} \right) {\cal 
L}_{(2)}^i\,, 
\ee
and 
\begin{eqnarray}
{\cal L}_{(2)}^1 &=& \nabla_a\nabla_b \phi \nabla^a\nabla^b \phi \,,\\
&&\nonumber\\
{\cal L}_{(2)}^2 &=& \left( \Box\phi\right)^2 \,,\\
&&\nonumber\\
{\cal L}_{(2)}^3 &=&  \left( \Box\phi\right)  
\nabla^a \phi \nabla^b \phi \nabla_a\nabla_b \phi \,,\\
&&\nonumber\\
{\cal L}_{(2)}^4 &=& \nabla^a \phi \nabla_b \phi \nabla_a\nabla_c \phi
\nabla^c\nabla^b \phi \,,\\
&&\nonumber\\
{\cal L}_{(2)}^5 &=& \left( \nabla^a \phi \nabla^b \phi \nabla_a\nabla_b 
\phi \right)^2 \,,
\end{eqnarray}
while the cubic terms are written as 
\be
{\cal A}_{(3)}^{abcdef} \nabla_a\nabla_b\phi \nabla_c\nabla_d\phi 
\nabla_e\nabla_f\phi =
\sum_{i=1}^{10} \, \beta_i \left( \phi, \bar{X} \right) {\cal L}_{(3)}^i 
\,,
\ee
where 
\begin{eqnarray}
{\cal L}_{(3)}^1 &=& \left( \Box\phi \right)^3 \,,\\
&&\nonumber\\
{\cal L}_{(3)}^2 &=&\left( \Box \phi \right) \nabla_a\nabla_b \phi  
\nabla^a\nabla^b \phi  \,,\\
&&\nonumber\\ 
{\cal L}_{(3)}^3 &=& \nabla_a\nabla_b \phi \nabla^b \nabla^c \phi \nabla^a 
\nabla_c\phi \,,\\
&&\nonumber\\
{\cal L}_{(3)}^4 &=& \left( \Box\phi \right)^2 \nabla_a\phi \nabla_b\phi 
\nabla^a\nabla^b \phi  \,,\\
&&\nonumber\\
{\cal L}_{(3)}^5 &=& \left( \Box\phi \right) \nabla_a\phi \nabla^c\phi 
\nabla^a\nabla^b \phi \nabla_b\nabla_c\phi \,,\\
&&\nonumber\\
{\cal L}_{(3)}^6 &=& \left( \nabla_a\nabla_b\phi \nabla^a\nabla^b\phi 
\right) \nabla_c\phi \nabla_d\phi \nabla^c\nabla^d \phi \,,\\
&&\nonumber\\
{\cal L}_{(3)}^7 &=& \nabla_a\phi \nabla_d\phi \nabla^a\nabla^b\phi 
\nabla_b\nabla_c\phi \nabla^c\nabla^d\phi \,,\\
&&\nonumber\\
{\cal L}_{(3)}^8 &=& \nabla_a\phi \nabla_c \phi \nabla_d\phi \nabla_e\phi 
\nabla^a\nabla^b\phi \nabla_b\nabla^c\phi 
\nabla^d\nabla^e\phi \,,\\
&&\nonumber\\
{\cal L}_{(3)}^9 &=& \left(\Box\phi \right) \left( \nabla_a\phi 
\nabla_b\phi 
\nabla^a\nabla^b \phi \right)^2 \,,\\
&&\nonumber\\
{\cal L}_{(3)}^{10} &=& \left( \nabla_a\phi \nabla_b \phi \nabla^a\nabla^b 
\phi \right)^3 \,.
\end{eqnarray}
The $\alpha_i\left( \phi, \bar{X} \right)$ and $\beta_i\left( \phi, 
\bar{X} \right)$ are arbitrary (regular) functions of their arguments,  
and so are the $ {\cal A}_{(2)}^{abcd} $ and $ {\cal 
A}_{(3)}^{abcdef} $. The requirement 
that the Ostrogradski instability is avoided imposes relations between 
these functions \cite{Langlois:2018dxi}.

Naturally, the search for exact solutions describing hairy black holes, 
stars, and other condensations has been a part of developing DHOST 
theories. Beginning with spherically symmetric solutions, and continuing 
with axially symmetric ones, the catalogue of analytical solutions is 
expanding, although slowly (see \cite{Faraoni:2021nhi} for a recent review 
of spherically symmetric solutions of GR and of general scalar-tensor 
gravity), including three-dimensional spacetimes \cite{Baake:2020tgk}. 
Most of the currently 
known solutions are stealth ones with a 
static geometry and a scalar field that depends linearly 
on time \cite{stealth1,stealth2,stealth3,stealth4, stealth5, stealth6, 
stealth7}. In Refs.~\cite{BenAchour:2018dap, Takahashi:2020hso, 
Charmousis:2019vnf}, general conditions for stealth solutions beyond shift 
symmetry have been obtained. However, the search for solutions beyond 
stealth ones has been 
started \cite{Nonstealth1, stealth3, stealth5}, the recent 
Refs.~\cite{BenAchour:2019fdf, Anson:2020trg, BenAchour:2020fgy}  leading 
the way in this 
effort. The novel approach used in \cite{BenAchour:2019fdf, Anson:2020trg,  
BenAchour:2020fgy}  is based on 
the use of disformal 
transformations \cite{Bekenstein:1992pj, Ezquiaga:2017ner, 
Zumalacarregui:2010wj, Zumalacarregui:2013pma} to generate new solutions 
from known ``seed'' ones; this technique is linked to the fact that the 
degeneracy classes of quadratic DHOST theories are stable under disformal 
transformations. Moreover, solutions of the coupled Einstein-Klein-Gordon 
equations are mapped into DHOST solutions and can be used as seeds.
Ref.~\cite{Achour:2021pla} discusses the Petrov classification of spacetimes in relation 
with  the construction of disformal solutions of DHOST theories and how 
the Petrov classes are mapped by disformal transformations. 
We refer to \cite{BenAchour:2019fdf, Achour:2021pla} and 
the references therein for the transformation properties of the many terms 
in the DHOST action under disformal transformations and their inverses.

In particular, Ref.~\cite{BenAchour:2019fdf} establishes the impossibility 
of ``disforming'' the Fisher-Janis-Newman-Winicour-Wyman solution of GR 
coupled with a free scalar field (which contains a naked singularity) 
\cite{Fisher:1948yn, Bergmann:1957zza, Janis:1968zz, Buchdahl:1972sj, 
Wyman:1981bd} into a black hole.\footnote{This solution is the general 
static, spherical, and asymptotically flat solution of the Einstein 
equations sourced by a free scalar field.}

Starting with a DHOST theory containing a metric tensor $g_{ab}$ and a 
scalar field $\phi$, a generic disformal transformation of the metric has 
the form \cite{Bekenstein:1992pj, Ezquiaga:2017ner, Zumalacarregui:2010wj, 
Zumalacarregui:2013pma}
\be
g_{ab} \rightarrow \tilde{g}_{ab}=\Omega^2 \left(\phi, X \right) g_{ab}+ 
W\left( \phi,  X\right) \nabla_a \phi \nabla_b \phi \,, \label{disformal}
\ee 
where $\Omega>0$ and $W$ are regular functions of the scalar field and of 
its gradient squared. The functions $\Omega(\phi,X)$ and $W(\phi,X)$ must 
satisfy the inequalities
\begin{eqnarray}
&& \Omega \neq 0 \,,  \label{Delta1}\\
&&\nonumber\\
&& \Omega^2 -X\left( \Omega^2\right)_{, X} -X^2 W_{,X} \neq 0 
\,,\label{Delta2}
\end{eqnarray}
to ensure that the map $g_{ab} \rightarrow \tilde{g}_{ab}$ 
is invertible \cite{BenAchour:2019fdf}.

When $W = 0$ the disformal transformation 
reduces to the usual conformal transformation (usually attributed to 
Bekenstein \cite{Bekenstein:1974sf}, but somehow anticipated in 
\cite{Higgs,Buchdahl72}) of ``old'' scalar-tensor gravity. This conformal 
transformation has been widely used as a solution-generating technique in 
this context ({\em e.g.}, \cite{Sneddon74,Abreuetal94, Yazadjiev:2001bx, 
Faraoni:2016ozb, Faraoni:2017afs,Faraoni:2017ecj, Chauvineau:2018zjy, 
Banijamali:2019gry} and references therein).

Here we want to establish, as much as is practically feasible, the nature 
of the image under disformal tranformations of spherically symmetric seeds 
describing, respectively, black holes, wormhole throats, or naked 
singularities (or, more in general, spherical solutions without horizons). 
In other words, we would like to know whether a black hole (or a 
whormhole, or a horizonless geometry) is mapped into another black hole, 
or a wormhole, or a horizonless solution.  The corresponding results for 
pure 
conformal transformations ({\em i.e.}, $W=0$) are reported in 
\cite{Faraoni:2015paa} (see also \cite{Hammad:2018ldj}); the discussion 
becomes significantly more complicated for disformal 
transformations~(\ref{disformal}). As exposed in the next section, it is 
possible to give a complete classification when starting with {\em static} 
seed solutions, but the analysis is not as conclusive (although some 
result can still be obtained) for stealth solutions in which $g_{ab}$ is 
static and the scalar $\phi$ has a linear dependence on time.

The authors of \cite{BenAchour:2019fdf} assess the presence 
or absence of 
black hole horizons by studing the norm squared of the Kodama vector, 
which is always defined in spherical symmetry, and changes sign on an 
(apparent) horizon. Our procedure of this section is equivalent but, in 
practice, 
streamlined and has the benefit of a more compact discussion. We use the 
fact that, in spherical symmetry, when (apparent) horizons exist they are 
located by the real and positive roots of the equation
\be
g^{ab}\nabla_a R \nabla_b R=0 \,,
\ee
where $R$ is the areal radius of the spherical geometry 
\cite{Misner:1964je,Abreu:2010ru, Faraoni:2015ula}.  A black hole horizon 
corresponds to a single real positive root while a double root identifies 
a wormhole throat, and no roots correspond to no horizons. We refer to 
apparent horizons because in time-dependent situations, which are of 
potential interest for DHOST solutions, horizons are not the null event 
horizons familiar from stationary black holes in GR, but they can be 
timelike or spacelike, or change their causal nature (see 
\cite{Nielsen:2008cr, Booth:2005qc, Faraoni:2015ula} for reviews). 
Apparent horizons have the drawback that they are foliation-dependent 
\cite{Wald:1991zz,Schnetter:2005ea}. However, in spherical symmetry, they 
become gauge-independent if one restricts oneself to spherically symmetric 
foliations \cite{Faraoni:2016xgy}, which will be assumed in the following.

As shown in the next sections, results can be established for static and 
stealth seed solutions, but the significant freedom of choosing the two 
functions $\Omega(\phi, X) $ and $W (\phi, X)$ allows for practically any 
outcome one desires when starting with general seeds and assigning   
these two functions. In Sec.~\ref{sec:3} and~\ref{sec:4} we provide two 
examples that are not static or stealth and show how to obtain new 
solutions of varied nature.

\section{Disformal mappings of static and of stealth solutions}
\label{sec:2}
\setcounter{equation}{0}

Consider a spherically symmetric seed solution which, without loss of 
generality, can be written in the form \cite{Weinberg}
\begin{align}
ds^2 &= g_{ab}dx^a dx^b = -A(t,r) dt^2 +\frac{dr^2}{B(t,r)} +r^2 
d\Omega_{(2)}^2 
\,,\label{seed}\\
\phi &=\phi(t,r) \,,
\end{align}
where $d\Omega_{(2)}^2 \equiv d\vartheta^2 +\sin^2 \vartheta \, 
d\varphi^2$ is the line element on the unit 2-sphere, $r$ is the areal 
radius, and $\phi$ is the scalar field acting as the only matter source. 
If this geometry possesses apparent horizons, they are located by 
the roots of the equation $\nabla^c r \nabla_c r =g^{rr}=B(t,r)=0$ 
\cite{Misner:1964je,Abreu:2010ru, Faraoni:2015ula}.

We use  the notation 
\be
\dot{\phi} \equiv \frac{\partial \phi (t,r)}{\partial t} \,, \quad \quad 
\phi' \equiv 
\frac{\partial \phi (t,r)}{\partial r} \,.
\ee
Under the disformal tranformation~(\ref{disformal}), the line 
element~(\ref{seed}) is mapped to
\be
d\tilde{s}^2 = -\left( \Omega^2 A-W\dot{\phi}^2 \right) dt^2 
+2W\dot{\phi}\phi' dtdr  + \left( \frac{\Omega^2}{B}  
+W\phi'^2\right) dr^2 + \tilde{R}^2 d\Omega_{(2)}^2 
\ee
and remains spherically symmetric, while the scalar field $\phi$ remains 
unchanged. Here 
\be
\tilde{R}(t,r)= \Omega \Big( \phi(t,r), X(t,r) \Big) r 
\ee
is the areal radius of the new metric $\tilde{g}_{ab}$.

\subsection{Static seed solutions}

Often a static and spherically symmetric solution, {\em i.e.}, one with $A=A(r), 
B=B(r), \phi=\phi(r)$  (for example, the Schwarzschild black 
hole \cite{BenAchour:2019fdf}) is used as  a seed. In this 
case  
$X(r) =B\phi'^2$ and the new line element, which is diagonal, is simply 
\be
d\tilde{s}^2 = -\Omega^2 A dt^2 +\left( \frac{\Omega^2}{B} 
+W\phi'^2\right) dr^2 + \tilde{R}^2 d\Omega_{(2)}^2 \,,
\ee
where $ \tilde{R}(r)=\Omega\Big( \phi(r), X(r) \Big) r$. We want to 
identify all the horizons of  this geometry,\footnote{One can, of course, 
transform to 
new coordinates using the 
areal radius $\tilde{R}$ as the radial coordinate, but this is unnecessary 
since 
eq.~(\ref{eq:stoAHs}) locating the apparent horizons (when they exist) is 
a scalar equation 
independent of the coordinate system.} which are located by the real 
positive roots of the equation \cite{Misner:1964je, Abreu:2010ru, 
Faraoni:2015ula} 
\be
\nabla^c \tilde{R} \nabla_c \tilde{R} =0   \label{eq:stoAHs}
\ee
(since these horizons are static, they are null event horizons). 
Since 
\be 
\nabla_{\mu} \tilde{R} =  \left\{ \left[ \Omega_{\phi} \phi' +\Omega_X 
\left( 
B\phi'^2\right)' 
\right]r +\Omega \right\} \delta_{\mu 1} \,,
\ee
eq.~(\ref{eq:stoAHs}) becomes 
\be
\frac{B}{ \Omega^2 +BW\phi'^2} \left[
 \Omega_{\phi} \phi' +\Omega_X \left( B\phi'^2\right)' +\frac{\Omega}{r} \right]^2 
= 0 \,.\label{questa}
\ee
The possible roots are all those of the  seed geometry, which are 
identified by $B=0$, 
plus the roots of  the new equation
\be
\Omega_{\phi} \phi' +\Omega_X \left( B\phi'^2\right)' +\frac{\Omega}{r} =0\,. 
\label{WHT}
\ee
As is clear from eq.~(\ref{questa}), if they exist, these new roots 
introduced by the disformal transformation are {\it always double roots 
associated with wormhole throats}. Therefore, the disformal transformation 
cannot generate new black hole horizons. In particular, this implies that:

\begin{itemize}

\item It is not possible to map a geometry without horizons into one with 
black hole horizons. The newly generated geometry can only have a wormhole 
throat or no horizons at all.

\item The disformal transformation cannot map a geometry characterized by a 
wormhole throat into a black hole.

\item It is possible that a black hole seed geometry is mapped into a 
wormhole throat. This can in principle happen when a new wormhole throat 
(a double root $\tilde{R}_{\rm WHT}$ of eq.~(\ref{WHT})) is introduced by 
the disformal transformation and the radius of the would-be black hole 
horizon $r_{\rm H}$, which corresponds to the areal radius $ 
\tilde{R}_{\rm H}=\Omega(\phi(r_{\rm H}), \phi'(r_{\rm H})) r_{\rm H}$, 
lies {\em below} the wormhole throat, {\em i.e.}, $\Omega(\phi(r_{\rm H})) 
r_{\rm H} < \tilde{R}_{\rm WHT}$.

\end{itemize}

Equation~(\ref{WHT}) can be written as 
\be
\frac{d\Omega}{dr} +\frac{\Omega}{r} =0 \,;
\ee
it is easy to see that this equation cannot be satified identically. In fact, its 
integration would give $ \Omega(r)= C/r$, where $C$ is an integration constant. This 
conformal factor is singular at $r=0$ and tends to zero as $r\rightarrow +\infty$, while 
$\Omega$ is instead required to be regular everywhere and go to unity as $r\rightarrow 
+\infty$ for asymptotically flat solutions.

We can discuss three cases separately.

\begin{enumerate}

\item If the seed solution has no horizons, there are no roots of 
$B(r)=0$, then there can only be a wormhole throat if real positive roots 
of eq.~(\ref{WHT}) exist. If no such throat exists, there are no horizons 
(in particular, if the seed solutions contains a naked singularity, it is 
mapped into another naked singularity spacetime).

\item If the seed solution  has  a black hole (event) horizon at 
$r=r_{\rm BH}$, then   
\be
B(r_{\rm BH})=0 \,, \quad\quad B'(r_{\rm BH}) >0 \,.
\ee
The equation locating the horizons in the new geometry is 
\be
B \left[ \Omega_{\phi} \phi' +\Omega_X \left( B\phi'^2\right)' 
+\frac{\Omega}{r} 
\right]^2 =0\,,
\ee
and all the roots $r_{\rm BH}$ are still roots but, as explained above, a 
new wormhole throat could be created by the disformal transformation, that 
relegates a black hole horizon to a region that is no longer part of the 
physical spacetime in the new geometry, if $\Omega(\phi(r_{\rm H})) r_{\rm 
H} < \tilde{R}_{\rm WHT}$.

\item Finally, suppose that the seed geometry has  a wormhole throat at 
$r_{\rm  W}$, where  $B(r_{\rm W})=B'(r_{\rm W})=0$. This remains a double 
root of 
\be
B \left[ \Omega_{\phi} \phi' +\Omega_X \left( B\phi'^2\right)' 
+\frac{\Omega}{r}  \right]^2 =0\,,
\ee
corresponding to the physical (areal) radius $ \tilde{R}_{\rm 
W}=\Omega\left( \phi(r_{\rm W}), X(r_{\rm W}) \right)  r_{\rm W}$.  If 
extra (real and positive) roots $ \tilde{R}_{\rm WW}$ exist, they are 
always double roots and they could obliterate the previous ones if $ 
\tilde{R}_{\rm W}< \tilde{R}_{\rm WW}$, in the manner described above for 
black hole horizons.

\end{enumerate}

\subsection{Stealth seed geometry}

Let us turn now to spherical stealth solutions 
\cite{stealth1,stealth2,stealth3,stealth4, stealth5, stealth6, stealth7} with
\be
A=A(r) \,, \quad\quad B=B(r) \,, \quad\quad \phi(t,r)=qt +\chi(r) \,,
\ee
where $q\neq 0 $ is a constant, giving $ X(r) = -\frac{q^2}{A}+ B\phi'^2 $, 
$\phi'=\chi'$, and the line element
\be
d\tilde{s}^2  = -\left( \Omega^2 A-q^2 W \right)  dt^2 +2qW \phi' dtdr + \left(  \frac{\Omega^2}{B} +W\phi'^2 \right) dr^2 + \tilde{R}^2 
d\Omega_{(2)}^2 \,,
\ee
with $ \tilde{R}(r)=\Omega\Big( \phi(r), X(r) \Big) r$. We now have 
\be
\nabla_{\mu} \tilde{R} = qr\Omega_{\phi} \delta_{\mu 0} + r \left[ \frac{\Omega}{r} + \Omega_{\phi}\phi' + \Omega_X \left( -\frac{q^2}{A} + B\phi'^2 
\right)'\right] 
\delta_{\mu 1}\,,
\ee
and the equation locating the apparent horizons is now 
\begin{align}
\tilde{g}^{ab}\nabla_a \tilde{R} \nabla_b \tilde{R} &= 
-q^2 \Omega_{\phi}^2 \left( \Omega^2 +BW\phi'^2\right) +2q^2 
\Omega_{\phi} BW\phi' \left[\frac{\Omega}{r} +\Omega_{\phi}\phi' 
+\Omega_X \left( -\frac{q^2}{A}+B\phi'^2\right)'\right] \nonumber\\ 
&\ \ \ \, + \left( \Omega^2A -q^2 W \right) B \left[ 
\frac{\Omega}{r} +\Omega_{\phi}\phi' +\Omega_X \left( -\frac{q^2}{A} 
+B\phi'^2\right)'\right]^2 =0 \,,\label{messy}
\end{align}
where we used the inverse metric 
\be
\left( \tilde{g}^{\mu\nu} \right) =  \left( \begin{array}{cccc}
( \frac{\frac{\Omega^2}{B}+W\phi'^2)}{g^{(2)}} & - 
\frac{qW\phi'}{g^{(2)}} & 0 & 0 \\
& & & \\
-\frac{q W\phi'}{g^{(2)}}  &  \frac{(q^2 W -\Omega^2 A)}{g^{(2)}}  & 0 
&0\\ 
& & & \\
0 & 0 &  \frac{1}{\tilde{R}^2} & 0 \\
& & & \\
0 & 0 & 0 & \frac{1}{ \tilde{R}^2 \sin^2 \vartheta} \\
\end{array} \right)\,,
\ee
and where 
\be
g^{(2)} = \Omega^2 \left( \frac{q^2 W}{B} -W A\phi'^2  -\frac{\Omega^2 A}{B} \right) <0\,,
\ee
is the determinant of the restriction of the seed metric $g_{ab}$ to the 
2-dimensional subspace $\left(t,r\right)$. In general, if they exist and 
they are real and positive, the roots of eq.~(\ref{messy}) are 
time-dependent due to the time-dependence of $\Omega, \Omega_{\phi}, 
\Omega_X$, and $W$. As a consequence, the corresponding horizons (when 
they exist) are dynamical apparent horizons (see, {\em e.g.}, 
\cite{Faraoni:2015ula}). The analysis of the roots of eq.~(\ref{messy}) 
(and, therefore, of the nature of the associated horizons) is more 
difficult and not much can be said in general without specifying the form 
of the functions $\Omega(\phi, X)$ and $W(\phi, X)$. We limit ourselves to 
the following considerations.

Suppose that the seed geometry has a black hole (event) horizon at 
$r_{\rm H}$, 
with $B(r_{\rm H})=0$ and $ B'(r_{\rm H})>0$: then eq.~(\ref{messy}) 
evaluated at $r_{\rm H}$ yields
\be 
\left( q  \Omega\Omega_{\phi} \right)^2 \Big|_{r_{\rm H} } =0 
\,.\label{eq:aba}
\ee 
Since $q\neq 0$ and $\Omega>0$, for the black hole 
horizon at $r_{\rm H}$ to  remain a (apparent) horizon (at $ 
\tilde{R}_{\rm 
H}=\Omega(r_{\rm H}) r_{\rm H}$) it is necessary that
\be
\Omega_{\phi}\left( \phi(t, r_{\rm H}), X( r_{\rm H}) \right)=0\,,
\label{mah?}
\ee
and, when this happens, we necessarily have a double root of 
eq.~(\ref{eq:aba}). Therefore, the (single root) black hole horizon of the 
``old'' spacetime is either eliminated by the disformal transformation or 
it is converted into a (double root) wormhole throat in the disformed 
world.

Due to the large amount of literature on shift-symmetric theories, it is 
relevant to restrict oneself to invertible disformal transformations and 
shift-symmetric potentials, as done in 
\cite{BenAchour:2019fdf}, in which case
\be
\Omega=\Omega(X)\,, \quad\quad W=W(X) \,.\label{questaquesta}
\ee
Under this assumption, eq.~(\ref{mah?}) is identically satisfied and a 
black hole horizon of the seed geometry is transformed into a wormhole 
throat. Moreover, eq.~(\ref{messy}) simplifies to
\be
\left( \Omega^2 A-q^2 W\right) B \left[ \frac{\Omega}{r} +\Omega_X \left( 
-\frac{q^2}{A}  +B\phi'^2 \right)'\right]^2=0 \,.\label{nonholeta}
\ee
In addition to the horizons of the seed metric (roots of $B=0$) already 
discussed, the additional potential roots given by the vanishing of the 
expression in square brackets $ \frac{\Omega}{r} +\Omega_X \left( 
-\frac{q^2}{A} +B\phi'^2 \right)'$ (if they exist and are real and 
positive), can only be double roots and, therefore, wormhole throats. The 
only possibility left to obtain a new root $ \tilde{R}_{\rm H}=r_{\rm H} 
\Omega\left( X(r_{\rm H}) \right)$ corresponds to the vanishing of the 
first bracket $\left( \Omega^2 A-q^2 W\right) $ in eq.~(\ref{nonholeta}), 
which happens when the functions appearing in the disformal 
transformation~(\ref{disformal}) satisfy the very specific relation
\be
W(X) -\frac{A(r)}{q^2} \, \Omega^2(X)=0 \quad \quad \mbox{at }\, r_{\rm 
H}=0\,.\label{nonni}
\ee
When it exists, this root is a black hole horizon (a single root) if the 
derivative with respect to $r$ of the left hand side of 
eq.~(\ref{nonni}) 
is non-vanishing, or
\be
X'\left( W_X-\frac{2A \Omega \Omega_X}{q^2}\right) - 
\frac{A'\Omega^2}{q^2} \neq 0  
\quad\quad \mbox{at }\, r_{\rm H} \,.
\ee
This possibility can be usefully exploited to generate new DHOST 
solutions by choosing the functions $\Omega(X)$ and $W(X)$ appropriately. 

When the seed solution is not static or stealth, the situation 
is more complicated. Due to the freedom of choosing the two functions 
$\Omega(\phi,X)$ and $W(\phi, X)$, the disformed universe can be endowed 
with black hole horizons even if the seed solution has none, as shown in 
the next section.

\section{Disforming Wyman's ``other'' solution}
\label{sec:3}
\setcounter{equation}{0}

Here we use as seed a little-known and simple solution of Einstein gravity 
with a free minimally coupled scalar field found long ago by Wyman 
\cite{Wyman:1981bd}.\footnote{Wyman's solution has been generalized to 
include a positive cosmological constant \cite{Sultana:2015lja} and has 
been used to generate a new family of dynamical solutions of Brans-Dicke 
gravity \cite{Banijamali:2019gry}.} The geometry and scalar field are
\begin{eqnarray} 
ds^{2} &=& -\kappa r^{2}dt^{2}+2dr^{2}+r^{2}d\Omega^{2}_{(2)} \,, 
\label{act} \\
&&\nonumber\\
\phi (t) &=& \phi_0  t \,, \label{phiWyman}
\end{eqnarray}
where $\kappa= 8\pi G$, $\phi_0$ is a dimensionless constant, and $r$ is 
the areal radius.  We refer to eqs.~(\ref{act}) and (\ref{phiWyman}) as 
Wyman's ``other'' solution to distinguish it from the better known 
Fisher-Janis-Newman-Winicour-Wyman solution 
\cite{Fisher:1948yn, Bergmann:1957zza, Janis:1968zz, 
Buchdahl:1972sj, Wyman:1981bd} which is the general solution of the 
Einstein equations sourced by a free scalar field with the properties of 
being  spherical, static, and asymptotically flat.

The geometry is static (but the scalar field is not) and it is not 
asymptotically flat: the spatial sections $t=$~const. are curved. This is 
not a stealth solution.  The Wyman geometry (\ref{act}) can also be 
recovered as a special case of a solution of scalar-tensor gravity found 
by Carloni and Dunsby \cite{Carloni:2013iip} which, however, has a scalar 
field that depends only on $r$, instead of $t$, and is subject to a 
power-law potential.

The Ricci scalar of the Wyman solution~(\ref{act}), (\ref{phiWyman}) 
\be
\mathcal{R} =\kappa \, g ^{ab}\, \nabla_a \phi\,\nabla_b \phi 
= -\frac{\phi_0^2 }{r^2}  \,, 
\ee
diverges as $r\rightarrow 0^{+}$. Since the equation $ g^{ab}\, \nabla_ a 
r \nabla_b r=0$ has no roots because $ g^{ab}\, \nabla_ a r \nabla_b r= 
g^{rr}= 1/2$, there are no horizons and the Wyman spacetime hosts a 
central 
naked singularity. The geometry is not asymptotically flat since the Ricci 
tensor
\be
{\cal R}_{ab}= \kappa \nabla_a \phi \nabla_b \phi 
= \kappa \phi_0^2 \delta_{a0} \delta_{b0} \,,
\ee
(which is independent of $r$) does not vanish as 
$r\rightarrow +\infty$. 
The Misner-Sharp-Hernandez mass of a sphere of radius $r$, defined by $ 
1-2M_\text{MSH}/r \equiv \nabla^{c} r \nabla_{c}r$ \cite{Misner:1964je, 
Hernandez:1966zia}, is $ M_\text{MSH}(r)= r/4 $.

Under the disformal transformation (\ref{disformal}), the Wyman  line 
element (\ref{act}) is mapped to 
\be
d\tilde{s}^2 = -\qty(\kappa r^2\Omega^2 - W\phi_0^2)dt^2 + 2\Omega^2 dr^2 
+ \tilde{R}^2 d\Omega_{(2)}^2   
\ee
(which is still diagonal), where $\tilde{R}(t,r) = \Omega \, r $. Since
\be
\nabla_{\mu} \tilde{R}= \delta_{\mu 0} \, \phi_0 \Omega_{\phi} r 
+\delta_{\mu 
1} \left( \Omega +\frac{2\phi_0^2}{\kappa \, r^2} \, \Omega_X \right)\,,
\ee
the equation $ \tilde{g}^{ab}\nabla_a \tilde{R} \nabla_b \tilde{R} =0$ 
locating the apparent horizons reads
\begin{eqnarray}
&&-2\phi_0^2 \Omega^2 \Omega_\phi^2 r^2 
+ \qty(\Omega + \frac{2\phi_0^2 \Omega_X}{\kappa r^2})^2\qty(\kappa 
r^2\Omega^2 - W\phi_0^2) = 0 \,.\nonumber\\
&& \label{lessmessy}  
\end{eqnarray}
We can now exploit the freedom given by $\Omega(\phi, X)$ and $W(\phi, X)$ 
to map the naked singularity into a black hole or a wormhole throat 
with corresponding static or dynamical  horizons. All the choices of 
$\Omega(\phi,X)$ and $W(\phi,X)$ in this section satisfy the 
inequalities~(\ref{Delta1}) and (\ref{Delta2}).

\subsection{$\Omega(X)=-X$}

First, we consider the conformal factor 
\be
\Omega(X)=-X =\frac{ \phi_0^2}{\kappa \, r^2}   \,; \label{OmegaX}
\ee
eq.~(\ref{lessmessy}) locating the apparent horizons becomes
\be
 \frac{\phi_0^6}{\kappa^2r^4}\qty(\frac{\phi_0^2}{\kappa r^2} - W) = 
0\,.\label{HoleOmegaX}
\ee
The new spacetime will still be horizonless if, for example
\begin{eqnarray}
W (X) & = & - X-\left( \frac{\phi_0^2}{\kappa \, X} \right)^2 = \frac{\phi_0^2}{\kappa r^2} - r^4\,,
\end{eqnarray}
because then eq.~(\ref{HoleOmegaX}) has no roots. 

Alternatively, the naked  singularity is mapped to a black hole with 
static apparent horizon if
\be
W (X) = \frac{r_0\sqrt{\kappa}}{\phi_0} \, \sqrt{\abs{X}} = \frac{r_0}{r}\,, 
\ee
(with $r_0$  a length scale), which yields the single root
\be
r_\text{H} = \frac{\phi_0^2}{\kappa r_0}\,,
\ee
corresponding to areal radius 
\be
\tilde{R}_\text{H} = \Omega_\text{H} \, r_\text{H} = r_0 \,.
\ee
Finally, the singular seed ends up as a wormhole throat with 
static 
apparent horizon if $ W=1 $, which generates the double root $ 
r_\text{H} = \phi_0 / \sqrt{\kappa} $ and 
\be
\tilde{R}_\text{H}  = \frac{\phi_0}{\sqrt{\kappa}} \,. 
\ee
Therefore, the central naked singularity of Wyman's ``other'' spacetime 
can be mapped to 
another horizonless geometry, or to a black hole with static apparent 
horizon, or to a static wormhole throat apparent horizon 
according to 
specific choices of the conformal and disformal factors $\Omega$ and $W$.

\subsection{$\Omega (X,\phi)=-\alpha X\phi$}

Another possible choice for the conformal factor is
\be
\Omega(X,\phi)=-\alpha X\phi  = \frac{\alpha \, \phi_0^3 \, t}{\kappa \, 
r^2}  \,, 
\ee
where $\alpha>0$ has the dimensions of an  inverse length. 
Eq.~(\ref{lessmessy}) becomes
\be
\frac{\alpha^2\phi_0^4}{\kappa}\qty(t^2 - \frac{2}{\kappa}) - Wr^2 = 0\,.
\ee
If we further choose $W = 1$, we find a double root corresponding to 
areal radius 
\be
\tilde{R}_\text{H} (t) = \frac{\phi_0t}{\sqrt{\kappa t^2 - 2  }}  
\ee
for $t>\sqrt{2/\kappa}$; it decreases with time and stabilizes as
\be
\tilde{R}_\text{H} (t) \xrightarrow[t\rightarrow \infty]{} 
\frac{\phi_0}{\sqrt{\kappa}}\,.
\ee

If we choose instead 
\be
W (X) =  \frac{ r_0 \sqrt{\kappa}}{\phi_0} \, \sqrt{-X} =  
\frac{r_0}{r} \,,
\ee
we find a single root corresponding to 
a dynamical black hole horizon with coordinate radius
\be
r_\text{H} (t) = \frac{\alpha^2\phi_0^4}{\kappa r_0}\qty(t^2 - 
\frac{2}{ \kappa})  
\ee
for $t>\sqrt{2/\kappa}$. This dynamical black hole apparent horizon in 
the new spacetime has physical radius
\be
\tilde{R}_\text{H} (t) = \frac{r_0 t}{\alpha\phi_0\qty(t^2 - 
2/\kappa )}\,,
\ee
which shrinks to zero as $t\rightarrow +\infty$.

\subsection{$ \Omega(\phi)=\alpha \phi$}

We finally consider a conformal factor of the form
\be
\Omega(\phi)=\alpha\phi =\alpha \, \phi_0 \, t  \,,
\ee
which determines the equation 
for the horizon radii 
\be
-2\alpha^2r^2 + \kappa r^2 \alpha^2 t^2 - W = 0 \,.
\ee
If 
\begin{eqnarray}
W(\phi,X) = -\frac{\alpha^2\phi^2}{X} - 
 \frac{\phi_0}{r_0\sqrt{\kappa}} \, \abs{X}^{-1/2} = \kappa\alpha^2 r^2 
t^2 - \frac{r}{r_0} \,,
\end{eqnarray}
one obtains the single root apparent horizon  radius $
 r_{\text{H}} = 1/( 2\alpha^2 r_0 )$  and the associated areal radius 
\be
\tilde{R}_\text{H}(t) = \frac{\phi_0 \, t}{2\alpha r_0} \label{Radius1}  
\ee
describing a linearly expanding black hole apparent horizon.

By choosing instead the constant disformal factor $W=\phi_0^2$, one finds 
a wormhole throat at radius 
\be
r_\text{H}(t)  = \frac{\phi_0}{\alpha\sqrt{\kappa t^2 - 2}}\,,
\ee
which is a double root at times $t> \sqrt{2/\kappa}$. Therefore, the 
areal radius of this horizon  
\be
\tilde{R}_\text{H}(t) = \alpha^2\, t  \sqrt{\kappa t^2 -2} 
\ee
expands. Coincidentally, the different choice 
\be
W (\phi) =  \frac{ \kappa \, \phi^2 -2\phi_0^2}{\left( 2\alpha \phi_0 
r_0\right)^2}  = \frac{\kappa t^2 - 2}{4\alpha^2r_0^2}  
\ee
leads to the same radius~(\ref{Radius1}) and  describes again a  
(double root) wormhole throat.

\section{Disforming the Husain-Martinez-Nu\~nez solution}
\label{sec:4}
\setcounter{equation}{0}

We can use another analytical solution of the Einstein equations found by 
Husain, Martinez and Nu\~nez \cite{Husain:1994} as a seed to disform. This 
seed solution describes a spherically symmetric geometry sourced by a free 
scalar field that is time-dependent. The geometry is also time-dependent, 
asymptotically Friedmann-Lema\^itre-Robertson-Walker (FLRW), and conformal 
to the Fisher-Janis-Newman-Winicour-Wyman solution. The 
Husain-Martinez-Nu\~nez line element in comoving time and the associated 
scalar field are \cite{Faraoni:2015ula}
\begin{align}
ds^2 = &-\qty(1 - \frac{2C}{r})^{\alpha} dt^2 + 
\frac{a^2(t)}{\qty(1-\frac{2C}{r})^{\alpha}}dr^2 
+a^2(t)r^2\qty(1-\frac{2C}{r})^{1-\alpha} d\Omega_{(2)}^2\,,\\
&\nonumber\\
\phi(t,r) =  &\pm \frac{1}{4\sqrt{\pi}}\ln\qty[D\qty(1 - 
\frac{2C}{r})^{\frac{\alpha}{\sqrt{3}}} a^{2\sqrt{3}}(t)]\,,
\end{align}
 where $\alpha = \pm\sqrt{3}/2$, $a(t) = a_0 t^{1/3}$, and $a_0, C$ and 
$D$  are non-negative constants. The areal radius is
\be
R(t) = a(t)r \qty(1-\frac{2C}{r})^{\frac{1 - \alpha}{2}} \,.
\ee
Applying the disformal transformation~(\ref{disformal}) to this geometry 
and using 
\be
\nabla_{\mu}\phi = \frac{\pm 1}{2\sqrt{3\pi}} \left[ 
\frac{\delta_{0\mu}}{t} 
+\frac{\alpha C}{r^2 \left( 1-2C/r\right)} \, \delta_{1\mu}   \right]  
\ee
gives
\begin{align}
d\tilde{s}^2 = &-\qty[\Omega^2\qty(1 - \frac{2C}{r})^{\alpha} - 
\frac{\overline{W}}{t^2}] dt^2 + \frac{2\alpha\overline{W} 
C}{tr^2\qty(1- 2C/r )} \,  dtdr
+ \qty[\frac{\Omega^2a_0^2  t^{2/3}}{\qty(1 - 2C/r)^{\alpha}} + 
\frac{\alpha^2 C^2 \overline{W}}{r^4\qty(1-2C/r)^2}]dr^2 \nonumber\\ 
&\nonumber\\
&+ \Omega^2a_0^2 \, t^{2/3} \, r^2 
\qty(1-\frac{2C}{r})^{1-\alpha}d\Omega_{(2)}^2 
\,,  
\end{align}
where $\overline{W} \equiv W/(12 \pi )$. The new areal radius is
\be
\tilde{R}(t) = \Omega \, ra_0 
t^{1/3}\qty(1-\frac{2C}{r})^{\frac{1-\alpha}{2}} \,.
\ee
If we choose 
\be
\Omega(t,r) = \frac{\qty(1- 2C/r)^{\alpha/2}}{a_0 \, t^{1/3}}  
\ee
(unfortunately, there is no explicit expression of $\Omega$ as  
a function of $\phi$ and $X$), the line element becomes
\begin{eqnarray}
d\tilde{s}^2 &=& 
-\qty[\frac{\qty(1-2C/r)^{2\alpha}}{  a_0^2\,t^{2/3}    } - 
\frac{\overline{W}}{t^2}]dt^2 + \frac{2\alpha C\overline{W}}{tr^2\qty(1 
- 2C/r)} dtdr + \qty[1+ \frac{\alpha^2C^2\overline{W}}{r^4\qty(1-2C/r)^2}]  
dr^2 \nonumber\\
&&\nonumber\\
&\, & + r^2\qty(1-\frac{2C}{r})d\Omega_{(2)}^2  \label{disformedHMN}
\end{eqnarray}
and is no longer conformal to the Fisher solution. 

The new areal radius is
\be
\tilde{R}(r) = r\sqrt{1-\frac{2C}{r}}  
\ee
and the equation locating the apparent horizons (when they exist)  reads
\begin{align}
\tilde{g}^{ab} \nabla_a \tilde{R} \nabla_b \tilde{R} &=  
\qty[\frac{\qty(1- 2C/r )^{2\alpha}}{a_0^2 \, t^{2/3}}  - 
\frac{\overline{W}}{t^2}]\times a_0^2 \,  
t^2r^4\qty(1-\frac{C}{r})^2\qty(1-\frac{2C}{r})^{1-2\alpha}
= 0 \,.
\end{align}
This equation is satisfied if $r_1 = C$, or $r_2 = 2C$, or if the term in 
square brackets vanishes. We discard the first root $r_1$ because 
$r_1<2C$ corresponds to imaginary $ \tilde{R}_1 $, while the second root 
is not a 
horizon since $r_2 = 2C$ corresponds to physical radius $\tilde{R}_2 = 0$.

The choice $W = 12\pi \, l_0^2$ with $l_0$ a length scale yields
\be
r_\text{H}(t) = \frac{2C \, 
t^{ \frac{2}{3\alpha}}}{ t^{\frac{2}{3\alpha}}-(l_0 a_0)^{\frac{1}{\alpha} 
}}\,, \label{aristadelsol}
\ee
with corresponding physical radius
\be
\tilde{R}_\text{H}(t) =  
\frac{2C \qty( l_0 a_0)^{ \frac{1}{2\alpha}}  t^{\frac{1}{3\alpha}} }{ t^{ 
\frac{2}{3\alpha}} -  \qty(l_0a_0)^{ \frac{1}{\alpha}} } \,.
\ee

For the parameter value $\alpha = \sqrt{3}/2$ of the original 
Husain-Martinez-Nu\~nez spacetime, the disformed metric is 
asymptotically flat and the horizon radius is 
\be
\tilde{R}_\text{H}(t) =  
\frac{2C \qty( l_0 a_0)^{ 1/ \sqrt{3} }  t^{\frac{2}{3\sqrt{3} }} }{ t^{ 
\frac{4}{3\sqrt{3}}} -  \qty(l_0a_0)^{ 2/ \sqrt{3} } } \,.\label{ABCDEF}
\ee
This radius is positive after a critical time $ t_* \equiv 
\qty(l_0a_0)^{3/2}$ (note that, since the scale factor $a(t)=a_0 t^{1/3}$ 
is dimensionless, the dimensions of $a_0$ are $\left[ a_0 \right] = \left[ 
L^{-1/3} \right]$ and $t_*$ has the dimensions of a time or length).

We can now deduce the history of the dynamical apparent horizon. Based on 
the existence of a single root one concludes that, when $0<t<t_*$, 
the disformed spacetime  has no horizons since $\tilde{R}_\text{H}<0$;   
a black hole apparent 
horizon begins to appear at $t=t_*$ with $ \tilde{R}_\text{H} 
(t_*)=+\infty$. 
It is always present at physical radius~(\ref{ABCDEF}) for $t>t_*$  and is 
approximated by  
\be 
\tilde{R}_\text{H}(t) \simeq 
\frac{2C\qty(l_0 a_0 )^{1/\sqrt{3}} }{t^\frac{2}{3\sqrt{3}}}\,, 
\ee
as $t \rightarrow +\infty$. This apparent horizon shrinks to zero at late 
times.

The other parameter value $\alpha = -\sqrt{3}/2$ gives 
\be
\tilde{R}_\text{H}(t) = \frac{ 2C \left( l_0a_0 \right)^{1/\sqrt{3} } \, 
t^{\frac{2}{3\sqrt{3} }}  }{ 
\left( l_0 a_0 \right)^{\frac{2}{\sqrt{3}} } -t^{ \frac{4}{3\sqrt{3}} } } 
\,.
\ee
Since the sign of the denominator is reversed with respect to 
eq.~(\ref{ABCDEF}), the horizon history is the time-reverse of that 
occurring for $\alpha=\sqrt{3}/2$.

The asymptotic behavior of the line element as $r\rightarrow 
+\infty$ (corresponding to $ \tilde{R} \rightarrow +\infty$) is 
\be
d\tilde{s}^2 \simeq -\qty( \frac{1}{a^2} - \frac{l_0^2}{t^2}) dt^2 + dr^2 
+ r^2d\Omega_{(2)}^2 \,.\label{Asyquesto}
\ee
We can redefine the time coordinate for this asymptotic metric according 
to
\be
dT\equiv \sqrt{\frac{1}{a_0^2 \, t^{2/3}} - \frac{l_0^2}{t^2}} \, dt \,,
\ee
which turns the asymptotic line element~(\ref{Asyquesto}) into  the 
Minkowski metric in coordinates $\left( \tau, r, \vartheta, \varphi 
\right)$, hence the new solution is asymptotically flat.

One can diagonalize the full line element~(\ref{disformedHMN}), 
obtaining (see Appendix~\ref{AppendixA}) 
\begin{eqnarray}
d\tilde{s}^2 &=& 
-F^2\left( \frac{A^{2\alpha} }{a^2} - \frac{l_0^2}{t^2}\right) dT^2 
+ \frac{\alpha^2 C^2 l_0^2}{A^{\alpha +1} \left(  1-C/r \right)^2} 
\qty[  \frac{A^2}{r^4} + \frac{A^{\alpha +2}}{ \alpha^2 C^2 l_0^2} - 
\frac{l_0^2}{ \left( A^{-\alpha} l_0^2 - t^2 A^{\alpha} /a^2 \right)r^4 }      
] d\tilde{R}^2 \nonumber\\
&&\nonumber\\
&\, & + \tilde{R}^2 d\Omega_{(2)}^2\,, \label{RedefMet}
\end{eqnarray}
where $F$ is an integrating factor and $ A \equiv 1-2C/r $. As 
$\tilde{R}\rightarrow+\infty$, $F\ \rightarrow 1$ and the line 
element~(\ref{RedefMet}) asymptotes to the Minkowski metric (see 
Appendix~\ref{AppendixA}).

\section{Conclusions}
\label{sec:5}
\setcounter{equation}{0}

Relatively few solutions of DHOST theories are known, even when symmetries 
such as spherical symmetry are imposed. Disforming a scalar field solution 
of Einstein theory selected as a seed constitutes a valuable 
solution-generating technique that has provided new analytical DHOST 
solutions and new insight into these theories, which are complicated and 
difficult to grasp. Most of the solutions generated using the disformal 
transformation used as seeds either static or stealth scalar field 
solutions of GR. We have discussed the disformal images of such seeds that 
describe black hole horizons, wormhole throats, or horizonless geometries. 
To move away from these rather limited situations, we have applied the 
disformal transformation to two lesser known {\em dynamical} scalar field 
solutions of Einstein theory, {\em i.e.}, Wyman's ``other'' solution and 
the Husain-Martinez-Nu\~nez geometry. The choices of conformal and 
disformal factors $\Omega$ and $W$ used were arbitrary, and designed to 
obtain horizons (single roots of eq.~(\ref{eq:stoAHs})), wormhole throats 
(double roots), or horizonless (no real positive roots) geometries. These 
examples demonstrate that pretty much any desired nature of the disformed 
solution with respect to horizons can be obtained thanks to the enormous 
freedom of choosing the functions $\Omega(\phi,X)$ and $W(\phi,X)$ 
arbitrarily.

Apart from seeds that are static or stealth solutions (considered here 
and, previously, in Ref.~\cite{BenAchour:2019fdf}), no prediction can be 
made on the result of the disformal transformation applied to arbitrary 
seeds, unless restrictions on $\Omega(\phi, X)$ and $W(\phi, X)$ are 
imposed. Due to the large number of free functions appearing in the DHOST 
action~(\ref{DHOSTaction}), it is not clear at present how to impose 
meaningful restrictions on $\Omega$ and $W$. Future research will 
hopefully provide some guidelines on how to choose these functions 
meaningfully from the physical point of view to restrict the scope of 
mathematical possibilities.

\bigskip
\appendix
\section{Diagonalization of the disformed Husain-Martinez-Nu\~nez line 
element~(\ref{disformedHMN})}
\label{AppendixA}
\renewcommand{\theequation}{A.\arabic{equation}}

Here we diagonalize the  metric~(\ref{disformedHMN}). We begin by using 
the areal radius 
\be
\tilde{R}(t,r)= r\sqrt{ 1-\frac{2C}{r}}\,, 
\ee
as the radial coordinate. Then, we replace $dr$ in 
eq.~(\ref{disformedHMN}) with
\be
dr= \frac{ \sqrt{ 1-2C/r}}{1-C/r} \, d\tilde{R} \,,
\ee
which leads to 
\begin{eqnarray}
d\tilde{s}^2 &=& -\left[ \frac{ \left(1-2C/r\right)^{2\alpha} }{a^2} 
-\frac{l_0^2}{t^2} \right] dt^2 + \frac{2l_0^2 \alpha C}{tr^2 \sqrt{1-2C/r} \left( 1-C/r\right)} \, 
dt d\tilde{R} \nonumber\\
&\nonumber\\
&\, & + \left[ 1+\frac{ \alpha^2 C^2 l_0^2}{r^4\left( 1-2C/r 
\right)^{\alpha} } 
\right] \frac{ \left(1-2C/r\right)}{ \left( 1-C/r \right)^2 } \, 
d\tilde{R}^2 + \tilde{R}^2 d\Omega_{(2)}^2 \,.
\end{eqnarray}
In order to eliminate the cross-term in $dtd\tilde{R}$, we redefine the 
time coordinate according to $t\rightarrow T$, with
\be
dT = \frac{1}{F} \qty(dt + \beta d\tilde{R})\,,
\ee
where $\beta \left( t,r \right)$ is a function to be determined 
and $F$ is an integrating factor guaranteeing that the differential 
$dT$ is exact and obeying the equation
\be
\pdv{}{\tilde{R}} \qty(\frac{1}{F}) = \pdv{}{t}\qty(\frac{\beta}{F})\,.
\ee
Using the new time, the line element becomes
\begin{eqnarray}
d\tilde{s}^2  &= &- F^2\qty(\frac{A^{2\alpha}}{a^2} - 
\frac{l_0^2}{t^2}) 
dT^2 + 2F\qty[\qty(\frac{A^{2\alpha}}{a^2}  - \frac{l_0^2}{t^2})\beta + 
\frac{\alpha C l_0^2}{tr^2\sqrt{A}\qty(1- C/r )}] 
dTd\tilde{R}\nonumber\\
&&\nonumber\\
&\, & +\qty[-\beta^2\qty(\frac{A^{2\alpha}}{a^2} - \frac{l_0^2}{t^2}) - 
2\beta \, \frac{\alpha C l_0^2}{tr^2 \sqrt{A}\qty(1- C/r )}
+ \frac{A}{\qty(1 - C/r )^2} \qty(1 + 
\frac{\alpha^2 C^2 l_0^2}{r^4 A^{\alpha} })]d\tilde{R}^2 \nonumber\\
&&\nonumber\\
&\, & + \tilde{R}^2d\Omega_{(2)}^2\,.
\end{eqnarray}
If we now set
\be
\beta (t, r) \equiv \frac{\alpha C l_0^2}{tr^2 
\sqrt{A}\qty(1-C/r )  \qty(\frac{l_0^2}{t^2} - 
\frac{A^{2\alpha}}{a^2})} \,,
\ee
the cross-term in $dt d\tilde{R}$ vanishes and we recover 
eq.~(\ref{RedefMet}).

As $\tilde{R} \rightarrow + \infty$, it is 
\be
\frac{\partial}{\partial 
\tilde{R}} \left( \frac{1}{F} \right) \simeq 0\,, 
\ee
asymptotically; then $F$ asymptotes to a constant and, by rescaling  
the 
time coordinate, it can be set to unity and $dt \simeq dT$. 
In this regime one  redefines the time coordinate 
$ t \simeq T \rightarrow \tau$ according to 
\be
d\tau = \sqrt{ \frac{1}{a^2} -\frac{ l_0^2}{t^2} } \, dT\,,
\ee
and the line element reduces to the Minkowski one in coordinates $\left( 
\tau, \tilde{R}, \vartheta, \varphi \right)$, 
\be
d\tilde{s}^2_{ \infty} \simeq -d\tau^2 
+d\tilde{R}^2 +\tilde{R}^2 d\Omega_{(2)}^2 \,.
\ee

\begin{acknowledgments} 

We thank a referee for useful comments. This work is supported, in part, 
by the Natural Sciences \& Engineering Research Council of Canada 
(Grant~2016-03803) and by Bishop's University.

\end{acknowledgments}


\begin{thebibliography}{99}

\bibitem{AmendolaTsujikawabook} L. Amendola and S. Tsujikawa, \emph{Dark 
Energy, Theory and Observations (Cambridge University Press)}, Cambridge, 
2010).

\bibitem{Brans:1961sx}
C.~Brans and R.~H.~Dicke,
\emph{Mach's principle and a relativistic theory of gravitation},
\emph{Phys. Rev.} \textbf{124}, 925-935 (1961)
doi:10.1103/PhysRev.124.925.

\bibitem{Bergmann:1968ve}
P.~G.~Bergmann,
\emph{Comments on the scalar tensor theory},
\emph{Int. J. Theor. Phys.} \textbf{1}, 25-36 (1968)
doi:10.1007/BF00668828.

\bibitem{Nordtvedt:1968qs}
K.~Nordtvedt,
\emph{Equivalence Principle for Massive Bodies. 2. Theory},
\emph{Phys. Rev. \textbf{169}}, 1017-1025 (1968)
doi:10.1103/PhysRev.169.1017.

\bibitem{Wagoner:1970vr}
R.~V.~Wagoner,
\emph{Scalar tensor theory and gravitational waves},
\emph{Phys. Rev. D} \textbf{1}, 3209-3216 (1970)
doi:10.1103/PhysRevD.1.3209.

\bibitem{Nordtvedt:1970uv}
K.~Nordtvedt, Jr.,
\emph{PostNewtonian metric for a general class of scalar tensor gravitational 
theories and observational consequences},
\emph{Astrophys. J.} \textbf{161}, 1059-1067 (1970)
doi:10.1086/150607.

\bibitem{Sotiriou:2008rp}
T.~P.~Sotiriou and V.~Faraoni, \emph{f(R) Theories of Gravity},
\emph{Rev. Mod. Phys.} \textbf{82}, 451-497 (2010)
doi:10.1103/RevModPhys.82.451
[arXiv:0805.1726 [gr-qc]].

\bibitem{DeFelice:2010aj}
A.~De Felice and S.~Tsujikawa, \emph{f(R) theories},
\emph{Living Rev. Rel.} \textbf{13}, 3 (2010)
doi:10.12942/lrr-2010-3
[arXiv:1002.4928 [gr-qc]].

\bibitem{Nojiri:2010wj}
S.~Nojiri and S.~D.~Odintsov,
\emph{Unified cosmic history in modified gravity: from F(R) theory to Lorentz 
non-invariant models}, \emph{Phys. Rept.} \textbf{505}, 59-144 (2011)
doi:10.1016/j.physrep.2011.04.001
[arXiv:1011.0544 [gr-qc]].

\bibitem{H1} C. Deffayet, G. Esposito-Far\`ese and A. Vikman, 
\emph{Covariant Galileon}, \emph{Phys. Rev.} D 79, 084003 (2009) 
arXiv:0901.1314.

\bibitem{H2} C. Deffayet, S. Deser and G. Esposito-Far\'ese,   
\emph{Generalized Galileons: All scalar models whose curved background 
extensions maintain second-order field equations and stress-tensors}, 
\emph{Phys. Rev. D} 80, 064015 (2009). arXiv:0906.1967.

\bibitem{H3} C. Deffayet, X. Gao, D. A. Steer and G. Zahariade, \emph{From 
k-essence to generalised Galileons}, \emph{Phys. Rev.} D 84, 064039 (2011). 
arXiv:1103.3260.

\bibitem{Horndeski} G. W. Horndeski, \emph{Second-order scalar-tensor field 
equations in a four-dimensional space}, \emph{Int. J. Theor. Phys.} 10, 363 
(1974). doi:10.1007/BF01807638.

\bibitem{GLPV1} J. Gleyzes, D. Langlois, F. Piazza and F. Vernizzi, 
\emph{Healthy theories beyond Horndeski}, \emph{Phys. Rev. Lett.} 114, no. 
21, 211101 (2015) arXiv:1404.6495.

\bibitem{GLPV2} J. Gleyzes, D. Langlois, F. Piazza and F. Vernizzi, 
\emph{Exploring gravitational theories beyond Horndeski}, \emph{JCAP} 1502, 018 (2015) 
arXiv:1408.1952.

\bibitem{DHOST1} D. Langlois and K. Noui, \emph{Degenerate higher derivative 
theories beyond Horndeski: evading the Ostrogradski instability}, \emph{JCAP} 
1602, no. 02, 034 (2016) arXiv:1510.06930.

\bibitem{DHOST2} D. Langlois and K. Noui, \emph{Hamiltonian analysis of higher 
derivative scalar-tensor theories}, \emph{JCAP} 1607, no. 07, 016 (2016) 
arXiv:1512.06820.

\bibitem{DHOST3} J. Ben Achour, D. Langlois and K. Noui, \emph{Degenerate 
higher order scalar-tensor theories beyond Horndeski and disformal 
transformations}, \emph{Phys. Rev. D} 93, no. 12, 124005 (2016) 
arXiv:1602.08398.

\bibitem{DHOST4} M. Crisostomi, K. Koyama and G. Tasinato, \emph{Extended 
Scalar-Tensor Theories of Gravity}, \emph{JCAP} 1604, no. 04, 044 (2016) 
arXiv:1602.03119.

\bibitem{DHOST5} H. Motohashi, K. Noui, T. Suyama, M. Yamaguchi and D. 
Langlois, \emph{Healthy degenerate theories with higher derivatives}, \emph{JCAP} 
1607, no. 07, 033 (2016) arXiv:1603.09355.

\bibitem{DHOST6} J. Ben Achour, M. Crisostomi, K. Koyama, D. Langlois, K. 
Noui and G. Tasinato, \emph{Degenerate higher order scalar-tensor theories 
beyond Horndeski up to cubic order}, \emph{JHEP} 1612, 100 (2016) 
arXiv:1608.08135.

\bibitem{DHOST7} M. Crisostomi, R. Klein and D. Roest, \emph{Higher 
Derivative Field Theories: Degeneracy Conditions and Classes}, \emph{JHEP} 
1706, 124 (2017) arXiv:1703.01623.

\bibitem{DHOSTreview1} D. Langlois, \emph{Dark energy and modiﬁed gravity 
in degenerate higher-order scalar-tensor (DHOST) theories: A review}, 
\emph{Int. J. Mod. Phys. D} 28, no. 05, 1942006 (2019) arXiv:1811.06271.

\bibitem{DHOSTreview2} D. Langlois, \emph{Degenerate Higher-Order 
Scalar-Tensor (DHOST) theories}, arXiv:1707.03625.

\bibitem{TheLIGOScientific:2017qsa} 
B.~P.~Abbott \textit{et al.} [LIGO Scientific and Virgo], \emph{GW170817: 
Observation of Gravitational Waves from a Binary Neutron Star Inspiral}, 
\emph{Phys. Rev. Lett.} \textbf{119}, no.16, 161101 (2017) 
doi:10.1103/PhysRevLett.119.161101 [arXiv:1710.05832 [gr-qc]].

\bibitem{Monitor:2017mdv} B.~P.~Abbott 
\textit{et al.} [LIGO Scientific, Virgo, Fermi-GBM and INTEGRAL], 
\emph{Gravitational Waves and Gamma-rays from a Binary Neutron Star Merger: 
GW170817 and GRB 170817A}, \emph{Astrophys. J. Lett.} \textbf{848}, no.2, L13 
(2017) doi:10.3847/2041-8213/aa920c [arXiv:1710.05834 [astro-ph.HE]].

\bibitem{Langloisetal18}D. Langlois, R. Saito, D. Yamauchi and K. Noui, 
\emph{Scalar-tensor theories and modiﬁed gravity in the wake of GW170817}, 
\emph{Phys. Rev. D 97}, no. 6, 061501 (2018) arXiv:1711.07403.

\bibitem{Creminellietal18} P. Creminelli, M. Lewandowski, G. Tambalo and 
F. Vernizzi, \emph{Gravitational Wave Decay into Dark Energy}, \emph{JCAP} 1812, no. 
12, 025 (2018) arXiv:1809.03484.

\bibitem{Wald:1984rg} R.~M.~Wald, \emph{General Relativity (Chicago 
University Press)}, Chicago, 1984). 
doi:10.7208/chicago/9780226870373.001.0001.

\bibitem{Langlois:2018dxi} D.~Langlois, 
\emph{Dark energy and modified gravity in degenerate higher-order 
scalar\textendash{}tensor (DHOST) theories: A review}, \emph{Int. J. Mod. 
Phys. D} \textbf{28}, no.05, 1942006 (2019) doi:10.1142/S0218271819420069 
[arXiv:1811.06271 [gr-qc]].

\bibitem{Faraoni:2021nhi} V.~Faraoni, A.~Giusti and B.~H.~Fahim, 
\emph{Spherical inhomogeneous solutions of Einstein and 
scalar\textendash{}tensor gravity: A map of the land},  Phys. Rept. 
\textbf{925}, 1-58 (2021) doi:10.1016/j.physrep.2021.04.003 
[arXiv:2101.00266 [gr-qc]].

\bibitem{Baake:2020tgk}
O.~Baake, M.~F.~Bravo Gaete and M.~Hassaine, \emph{Spinning black holes 
for generalized scalar tensor theories in three dimensions}, 
Phys. Rev. D \textbf{102}, no.2, 024088 (2020)
doi:10.1103/PhysRevD.102.024088
[arXiv:2005.10869 [hep-th]].

\bibitem{stealth1} E. Babichev and G. Esposito-Far\`ese, 
\emph{Time-Dependent Spherically Symmetric Covariant Galileons}, 
\emph{Phys. Rev. D} 87, 044032 (2013) arXiv:1212.1394.

\bibitem{stealth2} A. Anabalon, A. Cisterna and J. Oliva, 
\emph{Asymptotically locally AdS and flat black holes in Horndeski 
theory}, \emph{Phys. Rev. D} 89, 084050 (2014) arXiv:1312.3597.
 
\bibitem{stealth3} E. Babichev and C. Charmousis, \emph{Dressing a black 
hole with a time-dependent Galileon}, \emph{JHEP} 1408, 106 (2014) 
arXiv:1312.3204.

\bibitem{stealth4} C. Charmousis, T. Kolyvaris, E. Papantonopoulos and M. 
Tsoukalas, \emph{Black Holes in Bi-scalar Extensions of Horndeski Theories}, 
\emph{JHEP} 1407, 085 (2014) arXiv:1404.1024.

\bibitem{stealth5} T. Kobayashi and N. Tanahashi, \emph{Exact black hole 
solutions in shift symmetric scalar-tensor theories}, \emph{Prog. Theor. Exp. 
Phys.} 2014, 073E02 (2014) arXiv:1403.4364.
 
\bibitem{stealth6} E. Babichev and G. Esposito-Far\`ese, \emph{Cosmological 
self-tuning and local solutions in generalized Horndeski theories}, \emph{Phys. 
Rev. D} 95, no. 2, 024020 (2017) arXiv:1609.09798.

\bibitem{stealth7} H. Motohashi and M. Minamitsuji, \emph{General Relativity 
solutions in modiﬁed gravity}, \emph{Phys. Lett. B} 781, 728 (2018) 
arXiv:1804.01731.

\bibitem{BenAchour:2018dap} J.~Ben Achour and H.~Liu, \emph{Hairy 
Schwarzschild-(A)dS black hole solutions in degenerate higher order 
scalar-tensor theories beyond shift symmetry}, Phys. Rev. D \textbf{99}, 
no.6, 064042 (2019) doi:10.1103/PhysRevD.99.064042 [arXiv:1811.05369 
[gr-qc]].

\bibitem{Takahashi:2020hso} K.~Takahashi and H.~Motohashi, \emph{General 
Relativity solutions with stealth scalar hair in quadratic higher-order 
scalar-tensor theories}, JCAP \textbf{06}, 034 (2020) 
doi:10.1088/1475-7516/2020/06/034 [arXiv:2004.03883 [gr-qc]].

\bibitem{Charmousis:2019vnf} C.~Charmousis, M.~Crisostomi, R.~Gregory and 
N.~Stergioulas, \emph{Rotating Black Holes in Higher Order Gravity}, Phys. 
Rev. D \textbf{100}, no.8, 084020 (2019) doi:10.1103/PhysRevD.100.084020 
[arXiv:1903.05519 [hep-th]].

\bibitem{Nonstealth1} E. Babichev, C. Charmousis and A. Leh\'ebel, 
\emph{Asymptotically flat black holes in Horndeski theory and beyond}, \emph{JCAP} 
1704, no. 04, 027 (2017) arXiv:1702.01938.



\bibitem{BenAchour:2019fdf} J.~Ben Achour, H.~Liu and S.~Mukohyama, 
\emph{Hairy black holes in DHOST theories: Exploring disformal 
transformation as a solution-generating method}, \emph{JCAP} \textbf{02}, 
023 (2020) doi:10.1088/1475-7516/2020/02/023 [arXiv:1910.11017 [gr-qc]].

\bibitem{Anson:2020trg}
T.~Anson, E.~Babichev, C.~Charmousis and M.~Hassaine, 
``Disforming the Kerr metric,''
JHEP \textbf{01}, 018 (2021)
doi:10.1007/JHEP01(2021)018
[arXiv:2006.06461 [gr-qc]].

\bibitem{BenAchour:2020fgy} J.~Ben Achour, H.~Liu, H.~Motohashi, 
S.~Mukohyama and K.~Noui, \emph{On rotating black holes in DHOST 
theories}, JCAP \textbf{11}, 001 (2020) doi:10.1088/1475-7516/2020/11/001 
[arXiv:2006.07245 [gr-qc]].

\bibitem{Bekenstein:1992pj}
J.~D.~Bekenstein,
\emph{The Relation between physical and gravitational geometry},
\emph{Phys. Rev. D} \textbf{48}, 3641-3647 (1993)
doi:10.1103/PhysRevD.48.3641
[arXiv:gr-qc/9211017 [gr-qc]].

\bibitem{Ezquiaga:2017ner}
J.~M.~Ezquiaga, J.~Garc\'\i{}a-Bellido and M.~Zumalac\'arregui,
\emph{Field redefinitions in theories beyond Einstein gravity using the 
language of differential forms},
\emph{Phys. Rev. D} \textbf{95}, no.8, 084039 (2017)
doi:10.1103/PhysRevD.95.084039
[arXiv:1701.05476 [hep-th]].

\bibitem{Zumalacarregui:2010wj}
M.~Zumalacarregui, T.~S.~Koivisto, D.~F.~Mota and P.~Ruiz-Lapuente,
\emph{Disformal Scalar Fields and the Dark Sector of the Universe},
\emph{JCAP} \textbf{05}, 038 (2010)
doi:10.1088/1475-7516/2010/05/038
[arXiv:1004.2684 [astro-ph.CO]].

\bibitem{Zumalacarregui:2013pma} M.~Zumalac\'arregui and 
J.~Garc\'\i{}a-Bellido, \emph{Transforming gravity: from derivative 
couplings to matter to second-order scalar-tensor theories beyond the 
Horndeski Lagrangian}, \emph{Phys. Rev. D} \textbf{89}, 064046 (2014) 
doi:10.1103/PhysRevD.89.064046 [arXiv:1308.4685 [gr-qc]].

\bibitem{Achour:2021pla} J. Ben~Achour, A.~De Felice, 
M.A.~Gorji, S.~Mukohyama and M.~C.~Pookkillath, \emph{Disformal map and 
Petrov classification in modified gravity}, [arXiv:2107.02386 [gr-qc]].

\bibitem{Fisher:1948yn} I.~Z.~Fisher, \emph{Scalar mesostatic 
field with regard for gravitational effects}, \emph{Zh. Eksp. Teor. Fiz.} \textbf{18} 
(1948), 636-640 [arXiv:gr-qc/9911008 [gr-qc]].

\bibitem{Bergmann:1957zza} O.~Bergmann and R.~Leipnik,
\emph{Space-Time Structure of a Static Spherically Symmetric Scalar Field},
\emph{Phys. Rev.} \textbf{107} (1957), 1157-1161 doi:10.1103/PhysRev.107.1157.

\bibitem{Janis:1968zz} A.~I.~Janis, E.~T.~Newman and J.~Winicour, 
\emph{Reality of the Schwarzschild Singularity}, \emph{Phys. Rev. Lett.} 
\textbf{20} (1968), 878-880 doi:10.1103/PhysRevLett.20.878.

\bibitem{Buchdahl:1972sj} H.~A.~Buchdahl,
\emph{Static solutions of the Brans-Dicke equations},
\emph{Int. J. Theor. Phys.} \textbf{6} (1972), 407-412 doi:10.1007/BF01258735.

\bibitem{Wyman:1981bd} M.~Wyman, \emph{Static 
Spherically Symmetric Scalar Fields in General Relativity}, \emph{Phys. 
Rev. D} \textbf{24} (1981), 839-841 doi:10.1103/PhysRevD.24.839.

\bibitem{Bekenstein:1974sf}
J.~D.~Bekenstein,
\emph{Exact solutions of Einstein conformal scalar equations},
\emph{Ann. Phys. (N.Y.)}\textbf{82}, 535-547 (1974)
doi:10.1016/0003-4916(74)90124-9.

\bibitem{Higgs} P. Higgs, \emph{Quadratic Lagrangians and general 
relativity}, \emph{Nuovo Cimento} 11 (1959) 816–820.

\bibitem{Buchdahl72} H. A. Buchdahl, \emph{Static solutions of the 
Brans-Dicke equations}, \emph{Int. J. Theor. Phys.} 6 (1972) 407–412. 
doi:10.1007/BF01258735.

\bibitem{Sneddon74} G. E. Sneddon and C. B. G. McIntosh, \emph{Generation 
of solutions of the Brans-Dicke equations}, \emph{Aust. J.Phys.} 27, 
411-416 (1974) doi:10.1071/PH740411.

\bibitem{Abreuetal94} J. P. Abreu, P. Crawford, J. P. Mimoso, \emph{Exact conformal
scalar ﬁeld cosmologies},
\emph{Class. Quantum Grav.} 11 (1994) 1919–1940. arXiv:grqc/9401024,
doi:10.1088/0264-9381/11/8/002.

\bibitem{Yazadjiev:2001bx} S.~S.~Yazadjiev, \emph{Solution generating in 
scalar tensor theories with a massless scalar field and stiff perfect 
fluid as a source}, \emph{Phys. Rev. D} \textbf{65}, 084023 (2002) 
doi:10.1103/PhysRevD.65.084023 [arXiv:gr-qc/0108001 [gr-qc]].

\bibitem{Faraoni:2016ozb}
V.~Faraoni, F.~Hammad and S.~D.~Belknap-Keet,
\emph{Revisiting the Brans solutions of scalar-tensor gravity},
\emph{Phys. Rev. D} \textbf{94}, no.10, 104019 (2016)
doi:10.1103/PhysRevD.94.104019
[arXiv:1609.02783 [gr-qc]].

\bibitem{Faraoni:2017afs}
V.~Faraoni and S.~D.~Belknap-Keet,
\emph{New inhomogeneous universes in scalar-tensor and f(R) gravity},
\emph{Phys. Rev. D} \textbf{96}, no.4, 044040 (2017)
doi:10.1103/PhysRevD.96.044040
[arXiv:1705.05749 [gr-qc]].

\bibitem{Faraoni:2017ecj} V.~Faraoni, D.~K.~\c{C}iftci and 
S.~D.~Belknap-Keet, \emph{Symmetry of Brans-Dicke gravity as a novel 
solution-generating technique}, \emph{Phys. Rev. D} \textbf{97}, no.6, 
064004 (2018) doi:10.1103/PhysRevD.97.064004 [arXiv:1712.02205 [gr-qc]].

\bibitem{Chauvineau:2018zjy}
B.~Chauvineau,
\emph{New method to generate exact scalar-tensor solutions},
\emph{Phys. Rev. D} \textbf{100}, no.2, 024051 (2019)
doi:10.1103/PhysRevD.100.024051
[arXiv:1812.04934 [gr-qc]].

\bibitem{Banijamali:2019gry}
A.~Banijamali, B.~Fazlpour and V.~Faraoni,
\emph{Wyman\textquoteright{}s other scalar field solution,  
Sultana\textquoteright{}s
generalization, and their Brans-Dicke and $R^2$ relatives},
\emph{Phys. Rev. D} \textbf{100}, no.6, 064017 (2019)
doi:10.1103/PhysRevD.100.064017
[arXiv:1905.07023 [gr-qc]].

\bibitem{Faraoni:2016xgy} V.~Faraoni, G.~F.~R.~Ellis, J.~T.~Firouzjaee, 
A.~Helou and I.~Musco, \emph{Foliation dependence of black hole apparent 
horizons in spherical symmetry}, \emph{Phys. Rev. D} \textbf{95}, no.2, 
024008 (2017) doi:10.1103/PhysRevD.95.024008 [arXiv:1610.05822 [gr-qc]].

\bibitem{Wald:1991zz}
R.~M.~Wald and V.~Iyer,
\emph{Trapped surfaces in the Schwarzschild geometry and cosmic censorship},
\emph{Phys. Rev. D} \textbf{44}, R3719-R3722 (1991)
doi:10.1103/PhysRevD.44.R3719.

\bibitem{Schnetter:2005ea} E.~Schnetter and B.~Krishnan, 
\emph{Non-symmetric trapped surfaces in the Schwarzschild and Vaidya 
spacetimes}, \emph{Phys. Rev. D} \textbf{73}, 021502 (2006) 
doi:10.1103/PhysRevD.73.021502 [arXiv:gr-qc/0511017 [gr-qc]].

\bibitem{Nielsen:2008cr} A.~B.~Nielsen, \emph{Black holes and black hole 
thermodynamics without event horizons}, \emph{Gen. Rel. Grav.} 
\textbf{41}, 1539-1584 (2009) doi:10.1007/s10714-008-0739-9 
[arXiv:0809.3850 [hep-th]].

\bibitem{Booth:2005qc}
I.~Booth, \emph{Black hole boundaries},
\emph{Can. J. Phys.} \textbf{83}, 1073-1099 (2005)
doi:10.1139/p05-063
[arXiv:gr-qc/0508107 [gr-qc]].

\bibitem{Faraoni:2015paa}
V.~Faraoni, A.~Prain and A.~F.~Zambrano Moreno, \emph{Black holes and wormholes subject 
to conformal mappings}, \emph{Phys. Rev. D} \textbf{93}, no.2, 024005 (2016)
doi:10.1103/PhysRevD.93.024005
[arXiv:1509.04129 [gr-qc]].

\bibitem{Hammad:2018ldj} F.~Hammad, \emph{Revisiting black holes and 
wormholes under Weyl transformations}, \emph{Phys. Rev. D} \textbf{97}, 
no.12, 124015 (2018) doi:10.1103/PhysRevD.97.124015 [arXiv:1806.01388 
[gr-qc]].

\bibitem{Faraoni:2015ula}
V.~Faraoni, \emph{Cosmological and Black Hole Apparent Horizons}, 
\emph{Lect. Notes Phys.} \textbf{907} (Springer, New York, 2015).  
doi:10.1007/978-3-319-19240-6.

\bibitem{Weinberg} S. Weinberg, \emph{Gravitation and Cosmology: Principles and 
Applications of the General Theory of Relativity (Wiley, New York, 1972)}.


\bibitem{Abreu:2010ru} G.~Abreu and M.~Visser,
\emph{Kodama time: Geometrically preferred foliations of spherically symmetric 
spacetimes},
\emph{Phys. Rev. D} \textbf{82} (2010), 044027 doi:10.1103/PhysRevD.82.044027 
[arXiv:1004.1456 [gr-qc]].


\bibitem{Sultana:2015lja} J.~Sultana, \emph{Generating 
time dependent conformally coupled Einstein-scalar solutions}, \emph{Gen. Rel. 
Grav.} \textbf{47}, no.7, 73 (2015) doi:10.1007/s10714-015-1916-2.

\bibitem{Carloni:2013iip}
S.~Carloni and P.~K.~S.~Dunsby,
\emph{The 1+1+2 formalism for Scalar-Tensor gravity},
\emph{Gen. Rel. Grav.} \textbf{48}, no.10, 136 (2016)
doi:10.1007/s10714-016-2131-5
[arXiv:1306.2473 [gr-qc]].

\bibitem{Misner:1964je}
C.~W.~Misner and D.~H.~Sharp,
\emph{Relativistic equations for adiabatic, spherically symmetric 
gravitational collapse},
\emph{Phys. Rev.} \textbf{136}, B571-B576 (1964)
doi:10.1103/PhysRev.136.B571.

\bibitem{Hernandez:1966zia}
W.~C.~Hernandez and C.~W.~Misner,
\emph{Observer Time as a Coordinate in Relativistic Spherical 
Hydrodynamics},
\emph{Astrophys. J.} \textbf{143}, 452 (1966)
doi:10.1086/148525.

\bibitem{Husain:1994} V.~Husain, E.~A.~Martinez and D.~N\~unez, 
\emph{Exact solution for scalar field collapse}, \emph{Phys. Rev. D} 
\textbf{50}, 3783 (1994) doi:10.1103/PhysRevD.50.3783 [arXiv:gr-qc/9402021 
[gr-qc]].

\end{thebibliography}
\end{document}